\documentstyle [12pt] {article}

\newcommand{\cs}[3]{{{#3} \brace {#1 #2}}}

\newcommand{\h}[1]{\mathop{\lambda}\limits_{#1}\ \!\!\!}

\newcommand{\edf}{\ {\mathop{=}\limits^{\rm def}}\ }

\begin{document}
\begin{center}
\bf { TORSION ENERGY  }\\
\begin{center}
\bf{M.I.Wanas\footnote{Astronomy Department, Faculty of Science,
Cairo University, Giza, Egypt.

E-mail:wanas@mailer.eun.eg}}
\end{center}
\begin{abstract}
In the present work, torsion energy is defined. Its law of
conservation is given. It is shown that this type of energy gives
rise to a repulsive force which can be used to interpret
supernovae type Ia observations, and consequently the accelerating
expansion of the Universe. This interpretation is a pure geometric
one and is a direct application of the geometrization philosophy.
Torsion energy can also be used to solve other problems of General
Relativity especially the singularity problem.
\end{abstract}
\end{center}
\section {Introduction}
It is well known that gravity is one of the four fundamental
interactions, usually used to study and to interpret physical
phenomena. Although gravity is the weakest among these four
interactions, it is alone responsible for controlling the large
scale structure and evaluation of the Universe. Gravity has been
used, for a long time, to interpret most of large scale phenomenon
of our Universe. But recently, observations of supernovae (SN)type
Ia [1] show clearly that gravity, as we understand it today,
cannot account for such observations. These observations indicate
very clearly that the Universe is now in an accelerating expansion
phase. This implies the existence of a repulsive force which is
playing an important role in the structure and evolution of the
Universe. One can deduce that, either gravity is not well
understood and an important ingredient is missing in gravity
theories, or a new interaction is about to be discovered, in order
to account for such observations.

Gravity theories are usually constructed to give a better
understanding of gravitational interactions, from both
quantitative and qualitative points of view. Newton's theory deals
with gravity as a force acting at a distance. This theory has
suffered from some problems when applied to the motion in the
solar system (the advance of perihelion of Mercury's orbit). In
addition, the theory has shown its non-invariance under Lorentz
transformations. These problems motivated Einstein to construct a
new theory for gravity, the {\it "General theory of Relativity"}
(GR). In addition to the solution of the problems of Newton's
theory, GR gives very successful interpretation of gravitational
phenomena in the solar system, binary star system and many other
systems in the Universe. Also, GR has predicted the existence of a
number of phenomenon which has been confirmed observationally,
afterwards. Both Newton's theory and Einstein's theory give rise
to an attractive force. So, neither of these theories, in their
orthodox forms, can account for supernovae type Ia observations,
since this needs the existence of repulsive force which is missing
in both theories.

After the appearance of the problem of interpretation of SN
observations, many authors have returned back to a modified
version of GR, in which the cosmological constant existed in the
theory (cf. [2]).Although this constant can solve the problem,
since it gives rise to a repulsive force, it suffers from a big
problem the {\it " cosmological constant problem" } [3]. However,
GR still has its attractive feature i.e. its philosophical basis,
the {\it " geometrization philosophy"}. This philosophy still
deserves further investigations. It may provide solutions to
gravity problems especially those connected to SN observations.

The aim of the present work is to re-examine the geometrization
philosophy, seeking a solution for gravity problems.
\section{The Geometrization Philosophy}
In constructing his theory of GR, Einstein has invented an
ingenious idea that geometry can be used to solve physical
problems. This idea is known as the {\it " Geometrization of
Physics"}. It comprises a philosophical principle which can be
summarized in the following statement.

{\underline{\it "To understand Nature one has to start with
Geometry and end
with Physics "}}.\\
Einstein has applied this philosophy using the following guide
lines: \\
1. Laws of nature are identities in the chosen geometry. \\
2. Each physical quantity has a geometric representative. \\
3. Physical trajectories of test particles are geometric paths
(curves in the chosen

geometry). \\
Einstein has chosen the 4-dimensional Riemannian geometry with the
following identifications corresponding to the
3-guide lines given above [4]: \\
1. Conservation, as a law of nature, corresponds to the second
contracted Bianchi identity,
$$
 (R^{\mu \nu}- \frac{1}{2}g^{\mu \nu}R);_{\nu} \equiv 0   \eqno{(1)}
$$

where $R^{\mu \nu}$ is Ricci tensor, $R$ is Ricci scalar and the
semicolon is used for covariant

differentiation. \\
2. The conserved quantity (the quantity between brackets) ,
$$
G^{\mu \nu} \edf R^{\mu \nu}- \frac{1}{2} g^{\mu \nu}R, \eqno{(2)}
$$

is constructed from the curvature tensor (built using Christoffel
symbol). This

quantity corresponds to a type of energy that causes the curvature
of the space. For

this reason,  geometrically speaking, we are going to call the
quantity defined by (2)

{\bf "The
Curvature Energy"}.\\
3. Taking the third guide line into consideration, Einstein has
used the geodesic equation
$$
\frac{d^{2}x^{\mu}}{ds^{2}} +
\cs{\alpha}{\beta}{\mu}\frac{dx^{\alpha}}{ds}
\frac{dx^{\beta}}{ds} =0, \eqno{(3)}
$$

to represent the trajectory of any test particle in the field and
null geodesic

equation,
$$
\frac{d^{2}x^{\mu}}{d\lambda^{2}} +
\cs{\alpha}{\beta}{\mu}\frac{dx^{\mu}}{d\lambda}
\frac{dx^{\nu}}{d\lambda} =0, \eqno{(3)}
$$

to represent the trajectory of photons.\\ On the above scheme, we
have the following comments: \\
1. The choice of Riemannian geometry represents a special case,
since this geometry has

a vanishing torsion. This choice would not give a complete
description of the

physical world including space-time. Einstein has realized this
fact in his

subsequent attempts to construct unified field theories [4], [5]
and has used

geometries with torsion in these attempts. \\
2. The use of equation (3)or (4) to describe motion implies the
application of the

equivalence principle. This may represent a desirable feature in
describing the motion

of a scalar test particle (particle defined completely by its mass
(energy)). Note

that, most of elementary constituents of the Universe are fermions
(particles with

mass, spin, charge, .... ), for which (3) and (4)
are no longer applicable. \\
3. The most successful results of GR, which confirm the theory .
are those obtained using

the field equations for empty space (pure geometric). The use of
full field

equations of GR (equations containing a phenomenological matter tensor) is almost

problematic. \\
For these comments, we are going to examine the application of the
geometrization philosophy to geometries with torsion and
curvature.
 \section{Geometries with Torsion and Curvature}
Torsion tensor is the antisymmetric part of any non-symmetric
linear connection. Einstein has used two types of geometries, with
non-vanishing torsion, in his attempts to unify gravity and
electromagnetism. The first type is the Absolute Parallelism (AP)
geometry with non vanishing torsion but vanishing curvature [5].
The second  is of Riemann-Cartan type with simultaneously non-
vanishing torsion and curvature [4]. Calculations in first type
are more easier than in the second type. In what follows, we are
going to review very briefly a version of AP-geometry giving a
version in which both torsion and curvature are simultaneously non
vanishing. We have chosen this type since calculations in its
context are very easy.

The structure of a 4-dimensional space is defined completely by a
set of 4-contravariant linearly independent vector fields
$\h{i}^{\mu}$. The covariant components of such vector fields are
defined such that (for further details cf. [6]),
$$\h{i}^{\mu}\h{i}_{\nu}= \delta^{\mu}_{\nu}, \eqno{(5)}$$
$$\h{i}^{\mu}\h{j}_{\mu}= \delta_{ij} \eqno{(6)}.$$
The second order symmetric tensors,
$$g_{\mu \nu}  \edf \h{i}_{\mu}\h{i}_{\nu}, \eqno{(7)}$$
$$g^{\mu \nu}  \edf \h{i}^{\mu}\h{i}^{\nu}, \eqno{(8)}$$
can play the role of the metric tensor of the Riemannian space
associated with the AP-space. Using (7) and (8) one can define a
linear connection (Christoffel symbol of the second kind), using
which one can define covariant derivatives, as usual. Another
linear connection can be defined as a consequence of AP-condition,
$$
\h{i}_ {\stackrel{\mu}{+} | \nu}= 0 \eqno{(9)}
$$
which can be solved to give (the (+)sign is used to characterize
covariant derivative using the connection $\Gamma^{\alpha}_{. \mu
\nu}$)
$$
\Gamma^{\alpha}_{. \mu \nu} \edf \h{i}^{\alpha} \h{i}_{\mu , \nu}
= \cs{\mu}{\nu}{\alpha} + \gamma^{\alpha}_{. \mu \nu} \eqno{(10)}
$$
where $\cs{\mu}{\nu}{\alpha}$ is Christoffel symbol of second the
second kind and $\gamma^{\alpha}_{. \mu \nu}$ is the contortion
tensor, given by
$$
\gamma^{\alpha}_{. \mu \nu} \edf \h{i}^{\alpha} \h{i}_{\mu ; \nu}
\eqno{(11)}
$$
where the semicolon is used to characterize covariant
differentiation using Christoffel symbol. Now the torsion of the
space is defined by
$$
\Lambda^{\alpha}_{. \mu \nu} = \Gamma^{\alpha}_{. \mu \nu} -
\Gamma^{\alpha}_{. \nu \mu} = \gamma^{\alpha}_{\mu \nu} -
\gamma^{\alpha}_{\nu \mu }. \eqno{(12)}
$$
The relation between (11) and (12) can be written as [7],
$$
\gamma^{\alpha}_{\mu \nu} = \frac{1}{2} (\Lambda^{\alpha}_{.\mu
\nu}- \Lambda^{~~\alpha~~}_{\mu . \nu}- \Lambda^{~\alpha~}_{\nu
\mu}) \eqno{(13)}
$$
The relations(12) and (13) implies that the vanishing of the
torsion is a necessary and sufficient condition for the vanishing
of the contortion. The curvature tensor corresponding to the
linear connection (10) can be written, in the usual manner, as
$$
B^{\alpha}_{~\mu \nu \sigma} \edf \Gamma^{\alpha}_{~ \mu \sigma ,
\nu}- \Gamma^{\alpha}_{~ \mu \nu , \sigma} +
\Gamma^{\epsilon}_{~\mu \sigma}\Gamma^{\alpha}_{~\epsilon
\nu}-\Gamma^{\epsilon}_{~\mu \nu}\Gamma^{\alpha}_{~\epsilon
\sigma} . \eqno{(14)}
$$
This tensor vanishes identically because of (9). Although this
appears as a disappointing feature, but we are going to show that
it represents a cornerstone in the present work. The curvature
tensor (14) can be written in the form,
$$
B^{\alpha}_{~\mu \nu \sigma}= R^{\alpha}_{~\mu \nu \sigma}+
Q^{\alpha}_{~\mu \nu \sigma} \equiv 0,   \eqno{(15)}
$$
where,
$$
R^{\alpha}_{~\mu \nu \sigma} \edf \cs{\mu}{\sigma}{\alpha}_{, \nu}
- \cs{\mu}{\nu}{\alpha}_{, \sigma}+
\cs{\mu}{\sigma}{\epsilon}\cs{\epsilon}{\nu}{\alpha}
-\cs{\mu}{\nu}{\epsilon}\cs{\epsilon}{\sigma}{\alpha}, \eqno{(16)}
$$
is  the Riemann- Christoffel curvature tensor and the tensor
$Q^{\alpha}_{\mu \nu \sigma}$ is defined by
$$
Q^{\alpha}_{~\mu \nu \sigma} \edf \gamma^{\stackrel{\alpha}{+}}_{~
\stackrel{\mu}{+}\stackrel{\sigma}{+}|\nu}-\gamma^{\stackrel{\alpha}{+}}_{~
\stackrel{\mu}{+}\stackrel{\nu}{-}|\sigma}- \gamma^{\beta}_{~ \mu
\sigma}\gamma^{\alpha}_{\beta \nu} +  \gamma^{\beta}_{~ \mu
\nu}\gamma^{\alpha}_{\beta \sigma}, \eqno{(17)}
$$
where the (-)sign is used to characterize covariant derivatives
using the dual connection $\tilde \Gamma^{\alpha}_{. \mu \nu} (=
\Gamma^{\alpha}_{.\nu \mu} )$ .

Now to get a non-vanishing curvature, we have to parameterize (10)
by replacing it with [8]
$$
\nabla^{\alpha}_{. \mu \nu} = \cs{\mu}{\nu}{\alpha} + b ~
\gamma^{\alpha}_{. \mu \nu}, \eqno{(18)}
$$
where $b$ is a dimensionless parameter. The curvature tensor
corresponding to (18) can be written as,
$$
{\hat{B}}^{\alpha}_{.\mu \nu \sigma}= R^{\alpha}_{.\mu \nu
\sigma}+ b~ Q^{\alpha}_{.\mu \nu \sigma},   \eqno{(19)}
$$
which is, in general, a non vanishing tensor. The version of the
AP-geometry built using (18) is non as the PAP-geometry [9]. The
path equations corresponding to (18) can be written as
$$
\frac{d^{2}x^{\mu}}{d\tau^{2}} +
\cs{\alpha}{\beta}{\mu}\frac{dx^{\alpha}}{d\tau}
\frac{dx^{\beta}}{d\tau} = - b~~\Lambda^{..\mu}_{\alpha
\beta.}\frac{dx^{\alpha}}{d\tau} \frac{dx^{\beta}}{d\tau} .
\eqno{(20)}
$$
where $\tau$ is a scalar parameter. It can be easily shown that
(18) defines a non-symmetric, metric linear connection.
\section{Torsion Energy}
Using equation (15) we can deduce that
$$
 R^{\alpha}_{.\mu \nu \sigma}
\equiv - Q^{\alpha}_{.\mu \nu \sigma}.   \eqno{(21)}
$$
Although the tensors $R^{\alpha}_{. \mu \nu \sigma}$ and
$Q^{\alpha}_{. \mu \nu \sigma}$ appear to be mathematically
equivalent, they have the following differences:\\
1- The Riemann-Christoffel curvature tensor is made purely from
Christoffel symbols

(see (16)), while the tensor $Q^{\alpha}_{.\mu \nu \sigma}$ (17)
is made purely from the contortion (11)

( or from the torsion using (13)). The first tensor is
non-vanishing in Riemannian

geometry while the second vanishes in the same geometry. \\
2- The non-vanishing of $R^{\alpha}_{.\mu \nu \sigma}$ is the
measure of the curvature of the space, while the

addition of $Q^{\alpha}_{.\mu \nu \sigma}$ to it causes the space
to be flat. So, one is causing an inverse effect,

on the properties of space-time, compared to the other. For this
reason

we call $Q^{\alpha}_{.\mu \nu \sigma}$ "{\underline{\it{The
Curvature Inverse of Riemann-Christoffel Tensor}"}}. Note that

both tensors are considered as curvature tensor but one of them
cancels the effect of

the other.

Now, in view of the above two differences, we can deduce that
these two tensors are not, in general, equivalent. In other words,
if we consider gravity as curvature of space-time and is
represented by $R^{\alpha}_{. \mu \nu \sigma}$, we can consider
$Q^{\alpha}_{.\mu \nu \sigma}$ as representing
{\bf{anti-gravity}}! The existence of equal effects of gravity and
anti-gravity in the same system neutralizes the space-time,
geometrically. This situation is similar to the existence of equal
quantities of positive and negative electric charges in the same
system, which neutralizes the system electrically.

If we assume that gravity and anti-gravity effects are not equally
existing in the same system, then space-time curvature can be
represented by the tensor (19). The existence of anti-gravity
gives rise to a repulsive force, which can be used to interpret SN
type Ia observation. This can be achieved by adjusting the
parameter $b$ . The above discussion gives an argument on the
production of a repulsive force by torsion.

It is well known that the L.H.S. of (21) satisfies the second
Bianchi Identity, so using (21) we can easily show that

$$
\Sigma^{\alpha \beta}_{~~; \beta} \equiv 0 \eqno{(22)}
$$
where,
$$
\Sigma^{\alpha \beta} \edf Q^{\alpha \beta}- \frac{1}{2}g^{\alpha
\beta }Q , \eqno{(23)}
$$
$$
Q_{\alpha \beta} \edf Q^{\sigma}_{~. \alpha \beta \sigma}
\eqno{(24)}
$$
$$
Q \edf g^{\alpha \beta} Q_{\alpha \beta} . \eqno{(25)}
$$
It is clear from (22) that the physical quantity represented by
the tensor $ \Sigma^{\alpha \beta}$ is a conserved quantity. We
are going to call it the {\bf{"Torsion Energy"}}, since
$Q^{\alpha}_{.\mu \nu \sigma}$ is purely made of the torsion as
mentioned above (see (17)).

As a second argument on the existence of a repulsive force,
corresponding to the torsion of space-time, consider the
linearized form of (20) which can be written as [8],
$$
\Phi_{T} = \Phi_{N} (1-b)= \Phi_{N} + \Phi_{\Sigma}, \eqno{(26)}
$$
where,
$$
\Phi_{\Sigma} \edf - b \Phi_{N}, \eqno{(27)}
$$
 $\Phi_{N}$ is the Newtonian gravitational potential and
$\Phi_{T}$ is the total gravitational potential due the presence
of gravity and anti-gravity. It is clear from (26) that the
Newtonian potential is reduced by a factor $b$ due to the
existence of the torsion energy. It is obvious from (27) that $
\Phi_{\Sigma} $ and the Newtonian potential have opposite signs( $
b\geq 0,  [8]$ ). Then one can deduce that $\Phi_{\Sigma}$ is a
{\it repulsive gravitational potential}.
\section{Discussion and Concluding Remarks}
In the present work we have chosen a version of the 4-dimensional
AP-geometry to represent the physical world including space and
time. This geometry, PAP, is more wider than the Riemannian
geometry. The version of this geometry is characterized by the
parameter $b$. If $b = 0$ this geometry becomes Riemannian, while
if $b = 1$ it recovers AP-geometry.
We can draw the following remarks:\\

1- We have applied the geometrization philosophy and its guide
lines mentioned in section 2, to a geometry with curvature and
torsion. One can summarize the results of this application in the
following
points, corresponding to the three guide lines given in section 2. \\
(i) A conservation law, as a law of nature, giving conservation of
torsion energy is represented by the differential
identity (22). \\
(ii) The quantity called the {\it "torsion energy"} is represented
by the tensor (23), which is a part of a geometric structure. This
tensor
gives rise to anti-gravity and consequently a repulsive force. \\
(iii) Trajectories of test particles affected by the torsion of
the space-time is represented  by the path equation (20), which is
a curve in the geometric structure used.

The path equation (20) has been used to study trajectories of
spinning elementary test particles in a background with torsion
and curvature. The R.H.S. of this equation is suggested to
represent a type of interaction between torsion and the quantum
spin of the moving particle. The application of this equation to
the motion of the thermal neutrons in the Earth's gravitational
field removes the discrepancy in the COW-experiment [10]. \\

2- It is clear that {\it torsion energy}, defined in section 4.,
can solve the problem of SN type Ia observations, since it gives
rise to a repulsive force. This can be achieved by adjusting the
parameter $b$. One can now replace the term {\it{dark energy}} by
the term {\it{torsion energy}}. The later has a known origin (a
pure geometric origin). This shows the success of the
geometrization
scheme in dealing with physical problems. \\

3- Both terms {\it "dark energy} and the {\it "cosmological
constants"} are used as if they are metaphysical terms. They have
neither a geometric origin nor a well
defined physical origin.\\

4- Many authors consider $R^{\alpha}_{.\mu \nu \sigma}$and
$Q^{\alpha}_{.\mu \nu \sigma}$ as identical tensors from both
physical and mathematical points of view. For this reason, they
are using the term {\it{Teleparallel equivalent of GR}} for
theories built using $Q^{\alpha}_{.\mu \nu \sigma}$. In view of
the present work, such authors are, in fact, constructing
anti-gravity theories. But since gravitional and anti-gravitional
fields are phenomenologically equivalent, no discrepancy would
appear. Discrepancies would appear only if both fields exist in
the same
system, as in the case of SN type Ia observations. \\

5- In the context of the present work, we can extend the
geometrization
philosophy by adding the following postulate.\\
{\underline{\it{Physical phenomenae are
just interactions between the space-time structure and}}}\\
{\underline{\it{the intrinsic properties of the material constituents. }}} \\

6- Torsion energy can be used to give a reasonable solution for
the singularity problem in GR.
\section*{References}
{[1]} Tonry, J.L., Schmdit, B.P. et al. (2003) Astrophys. J.{\bf{594}}, 1. \\
{[2]} Mannheim, P.D. (2006) Prog.Part.Nucl.Phys. {\bf{56}} 340; gr-qc/0505266 \\
{[3]} Carroll, S.M. (2001) {\it"The Cosmological Constant"},
http://livingreviews.org/

Irr-2001-1. \\
{[4]} Einstein, A. (1955) {\it
"The Meaning of Relativity"
},Princeton, 5th ed.\\
{[5]} Einstein, A. (1930) Math. Annal. {\bf 102}, 685 .\\
{[6]} Wanas, M.I.  (2001) Cercet.Stiin.Ser.Mat. {\bf 10}, 297-309;
gr-qc/0209050.\\
{[7]} Hayashi, K. and Shirafuji, T. (1979) Phys.Rev. D{\bf 19},
3524. \\
{[8]} Wanas, M.I. (1998) Astrophys. Space Sci., {\bf{258}}, 237; gr-qc/9904019 \\
{[9]} Wanas, M.I. (2000) Turk. J. Phys. {\bf{24}}, 473; gr-qc/0010099 \\
{[10]} Wanas, M.I., Melek, M. and Kahil, M.E. (2000) Gravit.
Cosmol.,{\bf{6}}, 319.\\

\end{document}